\documentclass[conference,letterpaper]{IEEEtran}

\usepackage[utf8]{inputenc} 
\usepackage[T1]{fontenc}
\usepackage{url}
\usepackage{ifthen}
\usepackage{cite}
\usepackage[cmex10]{amsmath} % Use the [cmex10] option to ensure complicance
\usepackage[pdftex]{graphicx}
\usepackage{latexsym}
\usepackage{amssymb}
\usepackage{bm}
\usepackage{psfrag}
\usepackage{algorithm,algorithmic}

\newtheorem{definition}{Definition}
\newtheorem{theorem}{Theorem}
\newtheorem{lemma}{Lemma}

\newtheorem{corollary}{Corollary}
\newtheorem{remark}{Remark}

\newtheorem{example}{Example}

\newcommand{\vect}[1]{{\boldsymbol #1}}

\newcommand{\intg}[1]{\ensuremath{[\![#1]\!]} }
\newcommand{\1}{\mathbb{I}}

\newcommand{\bbZ}{\mathbb{Z}}

\newcommand{\dist}{\mathrm{d}}
\newcommand{\ot}{\leftarrow}

\begin{document}
\title{Distance Enumerators for Number-Theoretic Codes} 

%%% Single author, or several authors with same affiliation:
\author{%
  \IEEEauthorblockN{Takayuki Nozaki}
  \IEEEauthorblockA{
    Yamaguchi University, JAPAN\\
    Email: tnozaki@yamaguchi-u.ac.jp
  }
}

\maketitle

\begin{abstract}
  The number-theoretic codes are a class of codes defined by single or multiple congruences and are mainly used for correcting insertion and deletion errors.
  Since the number-theoretic codes are generally non-linear, the analysis method for such codes is not established enough.
  The distance enumerator of a code is a unary polynomial whose $i$th coefficient gives the number of the pairs of codewords with distance $i$.
  The distance enumerator gives the maximum likelihood decoding error probability of the code.
  This paper presents an identity of the distance enumerators for the number-theoretic codes.
  Moreover, as an example, we derive the Hamming distance enumerator for the Varshamov-Tenengolts (VT) codes.
\end{abstract}

\section{Introduction}
The number-theoretic codes \cite{helberg1993coding} are a class of codes defined by single or multiple congruences.
These codes are mainly used for correcting insertion and deletion errors \cite{varshamov1965code,levenshtein1966binary,bibak2018weight,nozaki2019bounded,nozaki2020weight}
or for correcting asymmetric errors \cite{varshamov1973class,shiozaki1982single}.
In general, the number-theoretic codes are non-linear.
Unfortunately, analysis methods for the number-theoretic codes have not been established enough compared with linear codes.

The distance enumerators for codes characterize the error correcting capability.
In particular, the Hamming distance enumerators \cite{delsarte1972bounds,kalai1995distance} are used for analyzing the error probability of the maximum likelihood decoder for the codes through symbol error channels.
The distance enumerator of a code is a unary polynomial whose $i$th coefficient gives the number of pairs of codewords with distance $i$.
For linear codes, we can easily derive the Hamming distance enumerator from the Hamming weight enumerator.
Thus, many works have investigated the Hamming distance enumerators for the linear codes.

On the other hand, for non-linear codes, the Hamming weight enumerators have been much researched.
Delsarte \cite{delsarte1972bounds} defined the Hamming distance enumerator for linear/non-linear codes
and derived an upper bound for the cardinality of code with designed Hamming distance.
Kalai and Linial \cite{kalai1995distance} analyzed the asymptotic behavior of growth rate for the Hamming distance enumerator.
Mimura \cite{mimura2015distance} analyzed the asymptotic growth rate for the Hamming distance enumerator for a class of non-linear code ensemble. 

We introduced the simultaneous congruence (SC) code, a general class of the number-theoretic code, in previous work \cite{nozaki2020weight}.
Moreover, we presented an identity for the Hamming weight enumerators for the SC codes and derived their cardinalities.
In this paper, we provide an identity for the distance enumerators for the SC codes.
This identity gives the distance enumerator related to not only Hamming distance but also other distances, e.g., Levenshtein distance \cite{levenshtein1966binary} and Lee distance \cite{lee1958some}.
This identity can be derived as a natural extension of the previous work.
This paper also derives the Hamming distance enumerators for Varshamov-Tenengoltz (VT) codes as an example.

The rest of the paper is organized as follows:
Section \ref{sec:def-nota} defines notations and definitions used throughout the paper.
Section \ref{sec:sc_wd} presents an identity for the distance enumerators for the SC codes.
Section \ref{sec:dd_VT} derives the Hamming distance enumerator for the VT codes and shows a numerical example.
Section \ref{sec:conc} concludes the paper.

\section{Definitions and Notations \label{sec:def-nota}}
This section gives the definitions and notations used throughout the paper.
Moreover, we define several codes and the distance enumerators.

Section \ref{ssec:note} defines the notations used throughout the paper.
Section \ref{ssec:codes} gives the two general classes of number-theoretic codes.
Section \ref{ssec:ewd} defines the several distances and the distance enumerators.

\subsection{Notations \label{ssec:note}}
Let $\mathbb{Z}$, $\mathbb{Z}^+$, $\mathbb{R}_{\ge 0}$, and $\mathbb{C}$ be the set of all integers, positive integers, non-negative real numbers, and complex numbers, respectively.
For $a,b\in\bbZ$, denote the integers between $a$ and $b$, by $\intg{a,b}$,
i.e., $\intg{a,b} := \{i\in\bbZ \mid a\le i \le b\}$.
In particular we denote $\intg{r} := \intg{0,r-1}$.
Let $\1\{P\}$ be the indicator function, which equals $1$ if the proposition $P$ is true and equals $0$ otherwise.
Denote the cardinality of a set $T$, by $|T|$.
Denote the vector of length $n$, by $\vect{x} = ( x_1, x_2,\dots, x_n )$.

For $a,b\in\mathbb{Z}$, we write $a \mid b$ if $a$ divides $b$.
For $a,b\in\mathbb{Z}$ and $n\in\mathbb{Z}^+$, denote $a \equiv b \pmod{n}$ if $(a-b)\mid n$.
For $a,b\in\mathbb{Z}$, let $\gcd(a,b)$ be the greatest common divisor of $a, b$.
Let $\mathrm{i}$ be the imaginary unit.
Define $e(x) := \exp (2\pi \mathrm{i} x)$.

\subsection{Number-Theoretic Codes \label{ssec:codes}}

Bibak and Milenkovic \cite{bibak2018weight} defined the binary linear congruence (BLC) codes as follows:
\begin{definition}[{\cite{bibak2018weight}}] \label{def:BLC}
  Denote the code length, by $n\in \mathbb{Z}^+$.
  Let $m\in\mathbb{Z}^+$, $\vect{h} = ( h_1, h_2, \dots, h_n ) \in\mathbb{Z}^n$, and $a\in\intg{m}$.
  Then, the BLC code of length $n$ with parameters $m, a, \vect{h}$ is defined by
    \begin{align*}
    \mathrm{BLC}_a(n,m,\vect{h}) 
    :=
    \{  &( x_1,x_2,\dots,x_n ) \in \{0,1\}^n \\
    &\mid \textstyle \sum_{i=1}^n h_i x_i \equiv a \pmod{m}  \}.
  \end{align*}
\end{definition}

We \cite{nozaki2020weight} defined the SC codes by extending the definition of the BLC codes as follows:
\begin{definition} [{\cite[Def.~2]{nozaki2020weight}}] \label{def:SC}
  Denote the code length, by $n\in\bbZ^+$.
  Let $r,s\in \bbZ^+$, $\vect{m} := ( m_1,m_2,\dots,m_s ) \in (\bbZ^{+})^s$, and 
  $\vect{a} := ( a_1,a_2,\dots,a_s ) \in \intg{m_1}\times\intg{m_2}\times \cdots \times\intg{m_s}$.
  For all $i\in \intg{1,s}$, let $\rho_i: \intg{r}^n \to \mathbb{Z}$ and
  denote $\vect{\rho} := ( \rho_1,\rho_2,\dots,\rho_s )$.
  Then, the $r$-ary SC code of length $n$ with parameters $s, \vect{\rho}, \vect{a}, \vect{m}$ is 
  \begin{align*}
    &C_{\vect{\rho},\vect{a},\vect{m}}(n,r,s) \\
    &:=
    \{ \vect{x}\in\intg{r}^n \mid
    \forall i\in\intg{1,s}~~
    \rho_i(\vect{x}) \equiv a_i \pmod{m_i}
    \}.
  \end{align*}
\end{definition}
There are examples of the number-theoretic codes included in the SC code in \cite{nozaki2020weight}.
For $\vect{h} = (h_1, h_2, \dots, h_n)\in \mathbb{Z}^n$, define a linear mapping $\ell_{\vect{h}}$ as $\ell_{\vect{h}}(\vect{x})=\sum_{i=1}^n h_i x_i$.
Then, the BLC codes is written as
\begin{equation*}
  C_{\ell_{\vect{h}},a,m}(n,2,1)
  =
  \mathrm{BLC}_a(n,m,\vect{h}).
\end{equation*}

Section \ref{sec:dd_VT} investigates Varshmov-Tenengoltz (VT) code \cite{varshamov1965code}, which is known as a single insertion/deletion correcting code.
Define a linear mapping $\omega:\intg{2}^n\to\mathbb{Z}$ as $\omega(\vect{x}) = \sum_{i=1}^n i x_i$.
Then, the VT code of length $n$ is defined by the following set:
\begin{equation*}
  \mathrm{VT}_a(n)
  :=
  \bigl\{ \vect{x} \in \{0,1\}^n \mid
  \omega(\vect{x}) \equiv a \pmod{n+1}
  \bigr\},
\end{equation*}
where $a\in\intg{0,n}$.
The cardinality of VT codes \cite{ginzburg1967certain}, \cite{stanley1972study} satisfies 
$|\mathrm{VT}_0(n)| \ge |\mathrm{VT}_a(n)|$ ($a \in \intg{0,n}$). 

\subsection{Distance and Distance Enumerator \label{ssec:ewd}}

A distance on $\intg{r}^n$ is a function $\dist:\intg{r}^n \times \intg{r}^n \to \mathbb{R}_{\ge 0}$, satisfying the following conditions:
\begin{enumerate}
\item For all $\vect{x},\vect{y}\in\intg{r}^n$, $\dist(\vect{x},\vect{y}) \ge 0$.
  Moreover, $\dist(\vect{x},\vect{y}) = 0$ iff $\vect{x} = \vect{y}$.
\item For all $\vect{x},\vect{y}\in\intg{r}^n$, $\dist(\vect{x},\vect{y}) = \dist(\vect{y},\vect{x})$.
\item For all $\vect{x},\vect{y},\vect{z}\in\intg{r}^n$, $\dist(\vect{x},\vect{y}) \le \dist(\vect{x},\vect{z}) + \dist(\vect{z},\vect{y})$.
\end{enumerate}

The following distances are often used in the coding theory.
\begin{enumerate}
\item
  The Hamming distance $\dist_{\mathrm{H}}(\vect{x},\vect{y})$ is the minimum number of substitutions required to change $\vect{x}$ to $\vect{y}$, i.e.,
  $\dist_{\mathrm{H}}(\vect{x},\vect{y}) := |\{ i\in\intg{1,n} \mid x_i \neq y_i \}|$.
\item 
  The longest common subsequence (or insdel) distance $\dist_{\mathrm{I}}(\vect{x},\vect{y})$ is the minimum number of insertions and deletions required to change $\vect{x}$ to $\vect{y}$.
\item 
  The Levenshtein distance $\dist_{\mathrm{Lev}}(\vect{x},\vect{y})$ \cite{levenshtein1966binary} is the minimum number of insertions, deletions, and substitutions required to change $\vect{x}$ to $\vect{y}$.
\item
  The Lee distance $\dist_{\mathrm{L}}(\vect{x},\vect{y})$ \cite{lee1958some} is defined by
  \begin{align*}
    &\dist_{\mathrm L}(\vect{x},\vect{y}) := \sum_{i=1}^n \delta_{\mathrm{L}} (x_i, y_i), \\
    &\delta_{L}(x,y) := \min\{ |x - y|, r - |x - y| \}.
  \end{align*}
\end{enumerate}

For a code $T\in \intg{r}^n$, the distance enumerator related to a distance $\dist$ is defined by
\begin{align*}
  \mathcal{D}(T;z)
  =
  \sum_{\vect{x}\in T} \sum_{\vect{y}\in T}z^{\dist (\vect{x},\vect{y})} 
  =
  \sum_{i=0}^{n} D_i z^{i}.
\end{align*}
where $D_i := |\{(\vect{x},\vect{y})\in T^2 \mid \dist(\vect{x},\vect{y}) = i\}|$ represents the number of pairs of codewords whose distance is $i$.
We denote the distance enumerator related to the Hamming distance $\dist_{\mathrm{H}}$ for a code $T$, by $\mathcal{D}_{\mathrm{H}}(T;z)$, and call it Hamming distance enumerator.

\begin{example} \label{exa:1}
  To simplify the notation, we denote the binary vector $(x_1,x_2,\dots, x_n)\in\intg{2}^n$, by $x_1x_2\dots x_n$.
  Consider $\mathrm{VT}_0(5) = \{00000, 10001, 01010, 00111, 11100, 11011 \}$.
  The Hamming distance enumerator for this code is
  \begin{equation*}
    \mathcal{D}_{\mathrm{H}}(\mathrm{VT}_0(5);z)
    =
    6z^0 + 8 z^2 + 16z^3 + 6z^4.
  \end{equation*}
\end{example}

\begin{remark}
  By normalizing the Hamming distance enumerator by the cardinality $|T|$ of the code,
  we get the average Hamming distance enumerator \cite{delsarte1972bounds,kalai1995distance} $\bar{\mathcal{D}}_{\mathrm{H}}(T;z)$.
  Since
  \begin{align*}
    \mathcal{D}_{\mathrm{H}}(T;0)
    =
    D_0
    =
    |\{(\vect{x},\vect{y})\in T^2 \mid \vect{x} = \vect{y}\}|
    =
    | T |,
  \end{align*}
  we get 
  \begin{align*}
    \bar{\mathcal{D}}_{\mathrm{H}}(T;z)
    :=
    \frac{\mathcal{D}_{\mathrm{H}}(T;z)}{|T|}
    =
    \frac{\mathcal{D}_{\mathrm{H}}(T;z)}{\mathcal{D}_{\mathrm{H}}(T;0)}  .
  \end{align*}
  In words, the average Hamming distance enumerator $\bar{\mathcal{D}}_{\mathrm{H}}(T;z)$ is derived from Hamming distance enumerator $\mathcal{D}_{\mathrm{H}}(T;z)$.
\end{remark}

This paper investigates the extended distance enumerator, which is a generalization of the distance enumerator.
\begin{definition} \label{def:ex-wt}
Let $n,r,s\in\bbZ^+$.
Let $\rho_i:\intg{r}^n\to\mathbb{Z}$ for $i\in\intg{1,s}$.
Denote $\vect{\rho} = (\rho_1,\rho_2,\dots,\rho_s)$, $\vect{u} = (u_1, u_2,\dots, u_{s})$, $\vect{v} = (v_1,v_1,\dots,v_{s})$.
We define the extended distance enumerator parameterized by $\vect{\rho}$ for a code $T\subseteq \intg{r}^n$ as 
\begin{align*}
  \mathcal{E}(T, \vect{\rho}; z, \vect{u},\vect{v})
  =
  \sum_{\vect{x}\in T}
  \sum_{\vect{y}\in T}
  z^{\dist(\vect{x},\vect{y})}
  \prod_{i\in \intg{1,s}} u_{i}^{\rho_i(\vect{x})} v_{i}^{\rho_i(\vect{y})} .
\end{align*}
In particular, we denote the extended Hamming distance enumerator, by $\mathcal{E}_{\mathrm{H}}(T, \vect{\rho}; z, \vect{u},\vect{v})$.
\end{definition}

\begin{example}
  We continue from Example \ref{exa:1}.
  The extended Hamming distance enumerator for $\mathrm{VT}_0(5)$ is
  \begin{align*}
    &\mathcal{E}_{\mathrm{H}}(\mathrm{VT}_0(5), \omega; z, u, v) \\
    &=
     \bigl(1 + u^6v^6\bigr)\bigl(1 + 2u^6v^6\bigr)
    +2\bigl(u^6 + v^6\bigr)\bigl(1 + u^{6}v^6 \bigr)z^2 \\
    &\quad+\bigl(1+u^6\bigr)\bigl(1+v^6\bigr)\bigl(u^6 + v^6 + 2u^6v^6\bigr) z^3 \\
    &\quad+\bigl(u^6+v^6\bigr)\bigl(u^6 + v^6 + u^6v^6\bigr) z^4.
  \end{align*}
\end{example}

\begin{remark} \label{rem:ele-wt}
  Define $\vect{1} := (1,1,\dots, 1)$.
  Then, the distance enumerator $\mathcal{D}(T;z)$ is derived from the extended distance enumerator $\mathcal{E}(T, \vect{\rho}; z, \vect{u}, \vect{v})$ as follows:
  \begin{equation*}
    \textstyle
    \mathcal{E}(T, \vect{\rho}; z,\vect{1},\vect{1})
    =
    \mathcal{D}(T; z) .
  \end{equation*}
\end{remark}

\section{Extended Distance Enumerators for SC Codes \label{sec:sc_wd}}

\subsection{Main Result and Corollary}
The following theorem presents an important formula to derive the extended distance enumerator.
\begin{theorem} \label{the:SC-WT}
  Define the SC codes (resp.\ extended weight enumerator) as in Definition \ref{def:SC} (resp.\ \ref{def:ex-wt}).
  Define $\vect{u} e\bigl(\frac{\vect{k}}{\vect{m}}\bigr)
  := \bigl(u_1 e\bigl(\frac{k_1}{m_1}\bigr), u_2e\bigl(\frac{k_2}{m_2}\bigr),\dots, u_se\bigl(\frac{k_s}{m_s}\bigr)\bigr)$.
  Then the following identity holds:
  \begin{align}
    &\mathcal{E}(C_{\vect{\rho},\vect{a},\vect{m}}(n,r,s), \vect{\rho}; z, \vect{u},\vect{v}) \notag \\
    &=
    \sum_{j_1,k_1\in\intg{m_1}}\sum_{j_2, k_2\in\intg{m_2}} \cdots \sum_{j_s, k_s\in\intg{m_s}} \notag \\
    &\hspace{10mm}\mathcal{E} \bigl(\intg{r}^n, \vect{\rho}; z, \vect{u}e\bigl(\tfrac{\vect{j}}{\vect{m}}\bigr), \vect{v}e\bigl(\tfrac{\vect{k}}{\vect{m}}\bigr)\bigr) \notag \\
    &\hspace{20mm}\times\prod_{i=1}^{s} \frac{1}{m_i^2}e \biggl(- \frac{a_i(j_i+k_i)}{m_i}\biggr).
    \label{eq:ex-dd-id}
  \end{align}
\end{theorem}

\begin{remark}
  In some special cases, we are able to obtain an explicit formula of the extended distance enumerator $\mathcal{E} \left(\intg{r}^n, \vect{\rho}; z, \vect{u}, \vect{v}\right)$.
  In such cases, Theorem \ref{the:SC-WT} presents the extended distance enumerator for the SC code $C_{\vect{\rho},\vect{a},\vect{m}}(n,r,s)$.
\end{remark}

The following corollary gives the Hamming distance enumerators for the BLC codes.
\begin{corollary} \label{cor:BLC}
  Define the parameters of BLC codes as in Definition \ref{def:BLC}.
  Then, the Hamming distance enumerators for the BLC codes are 
  \begin{align*}
    &\mathcal{E}_{\mathrm{H}}(\mathrm{BLC}_{a}(n,m,\vect{h}); z)
    \\ &=
    \sum_{j,k\in\intg{m}} 
    \frac{1}{m^2}e \biggl(- \frac{a(j+k)}{m}\biggr)
    \notag \\
    &\quad\times
    \prod_{i=1}^{n} \biggl( 1 + e\biggl(\frac{h_i j}{m}\biggr)z
    + e\biggl(\frac{h_i k}{m}\biggr) z + e\biggl(\frac{h_i (j+ k)}{m}\biggr) \biggr) . 
  \end{align*}
\end{corollary}

\subsection{Proofs}
The proof of Theorem \ref{the:SC-WT} uses the following identity for the code membership function. 
\begin{lemma} [{\cite[Eq.~(4)]{nozaki2020weight}}] \label{lem:indi}
  Define the parameters of SC codes as in Definition \ref{def:SC}.
  The following identity holds for any $\vect{\rho}, \vect{a}, \vect{m}$:
  \begin{align*}
  &\1\{\vect{x}\in C_{\vect{\rho},\vect{a},\vect{m}}(n,r,s)\}
  \\ &=  
  \prod_{i=1}^{s} \sum_{k_i\in\intg{m_i}} \frac{1}{m_i}
  e\biggl( \frac{- a_ik_i}{m_i} \biggr)
  \biggl( e\biggl( \frac{ k_i}{m_i} \biggr) \biggr)^{\rho_i(\vect{x})}  .
  \label{eq:pr-SC-WT1}
  \end{align*}
\end{lemma}

\subsubsection{Proof of Theorem \ref{the:SC-WT}}
Combining Lemma \ref{lem:indi} and Definition \ref{def:ex-wt}, we get 
\begin{align*}
  &\mathcal{E}(C_{\vect{\rho},\vect{a},\vect{m}}(n,r,s), \vect{\rho}; z, \vect{u},\vect{v}) 
  \\  &=
  \sum_{\vect{x}\in \intg{r}^n}
  \sum_{\vect{y}\in \intg{r}^n}
  z^{\dist(\vect{x},\vect{y})}
  \prod_{i = 1}^{s} u_{i}^{\rho_i(\vect{x})} v_{i}^{\rho_i(\vect{y})} \\
  &\qquad\times \1\{\vect{x}\in C_{\vect{\rho},\vect{a},\vect{m}}(n,r,s)\}
   \1\{\vect{y}\in C_{\vect{\rho},\vect{a},\vect{m}}(n,r,s)\}
   \\  &=
  \sum_{\vect{x}\in \intg{r}^n}
  \sum_{\vect{y}\in \intg{r}^n}
  z^{\dist(\vect{x},\vect{y})}\prod_{i=1}^{s} \sum_{j_i, k_i\in\intg{m_i}} \frac{1}{m_i^2}
  e\biggl( - \frac{ a_i (j_i+k_i) }{m_i} \biggr)
   \\
  &\qquad\times  
  \biggl( e\biggl( \frac{ j_i}{m_i} \biggr) u_i \biggr)^{\rho_i(\vect{x})}
  \biggl( e\biggl( \frac{ k_i}{m_i} \biggr) v_i \biggr)^{\rho_i(\vect{y})}
   \\  &=
   \sum_{j_1, k_1\in\intg{m_1}}   \sum_{j_2, k_2\in\intg{m_2}} \!\cdots\! \sum_{j_s, k_s\in\intg{m_s}}
   \Biggl[ \prod_{i=1}^{s} \frac{1}{m_i^2}  e\Bigl( {\textstyle - \frac{ a_i (j_i+k_i) }{m_i} } \Bigr) \Biggr]
   \\
  &\quad\times  
   \sum_{\vect{x}, \vect{y}\in \intg{r}^n}
   z^{\dist(\vect{x},\vect{y})}
   \prod_{i=1}^{s}
   \bigl( e\bigl( {\textstyle \frac{ j_i}{m_i} } \bigr) u_i \bigr)^{\rho_i(\vect{x})}
   \bigl( e\bigl( {\textstyle \frac{ k_i}{m_i} } \bigr) v_i \bigr)^{\rho_i(\vect{y})}
   \\  &=
   \sum_{j_1, k_1\in\intg{m_1}}   \sum_{j_2, k_2\in\intg{m_2}} \!\cdots\! \sum_{j_s, k_s\in\intg{m_s}} 
   \Biggl[ \prod_{i=1}^{s} \frac{1}{m_i^2}  e\Bigl( {\textstyle - \frac{ a_i (j_i+k_i) }{m_i} } \Bigr) \Biggr]
   \\
  &\qquad\times  
   \mathcal{E}(\intg{r}^n, \vect{\rho}; z, \vect{u}e(\vect{j}/\vect{m}),\vect{v}e(\vect{k}/\vect{m})).
\end{align*}

\subsubsection{Proof of Corollary \ref{cor:BLC}}
The following holds:
\begin{equation}
  \mathcal{E}_{\mathrm{H}} \bigl(\intg{2}^n, \ell_{\vect{h}}; z, u, v\bigr)
  =
  \prod_{i=1}^{n} \bigl\{ 1 + u^{h_i}z + v^{h_i}z + (uv)^{h_i} \bigr\}.
  \label{eq:DD_whole_BLC}
\end{equation}
Combining this identity and \eqref{eq:ex-dd-id}, we get
\begin{align*}
  &\mathcal{E}_{\mathrm{H}} (\mathrm{BLC}_{a}(n,m,\vect{h}), \ell_{\vect{h}}; z, u,v) \\
  &=
  \sum_{j,k\in\intg{m}} 
  \frac{1}{m^2}e \biggl(- \frac{a(j+k)}{m}\biggr)
  \prod_{i=1}^{n} \biggl( 1 + u^{h_i} e\biggl(\frac{h_i j}{m}\biggr)z
  \notag \\
  &\qquad\qquad+ v^{h_i} e\biggl(\frac{h_i k}{m}\biggr) z + (u v)^{h_i}  e\biggl(\frac{h_i (j+ k)}{m}\biggr) \biggr)  
\end{align*}
From this equation and Remark \ref{rem:ele-wt}, we have
\begin{align*}
  &\mathcal{D}_{\mathrm{H}}(\mathrm{BLC}_{a}(n,m,\vect{h}); z)
  \\ &=
  \mathcal{E}_{\mathrm{H}}(\mathrm{BLC}_{a}(n,r,m), \ell_{\vect{h}}; z, 1, 1)
  \\ &=
  \sum_{j,k\in\intg{m}} 
  \frac{1}{m^2}e \biggl(- \frac{a(j+k)}{m}\biggr)
  \notag \\
  &\quad\times
  \prod_{i=1}^{n} \biggl( 1 + e\biggl(\frac{h_i j}{m}\biggr)z
  + e\biggl(\frac{h_i k}{m}\biggr) z + e\biggl(\frac{h_i (j+ k)}{m}\biggr) \biggr).  
\end{align*}

\section{Hamming Distance Enumerator for VT Codes \label{sec:dd_VT}}

This section derives the Hamming distance enumerator for the VT codes.
Section \ref{ssec:VT-prop} gives some properties which are useful to calculate the Hamming distance enumerator.
Section \ref{ssec:VT-comp} shows algorithms to calculate the Hamming distance enumerator efficiently.
Section \ref{ssec:VT-exam} gives a numerical example.

\subsection{Properties \label{ssec:VT-prop}}
To simplify the notation, we denote $m:= n+1$.
Corollary \ref{cor:BLC} leads the Hamming distance enumerators for the VT codes:
\begin{align}
  &\mathcal{D}_{\mathrm{H}}(\mathrm{VT}_{a}(n); z)
  =
  \frac{1}{m^2}
  \sum_{j,k\in\intg{m}} e \biggl(- \frac{a(j+k)}{m}\biggr) A_{m,j,k}(z),
    \notag \\ 
  &A_{m,j,k}(z)
    :=
    \prod_{i=1}^{m-1}
    \Bigl( \textstyle 1 + e\Bigl(\frac{i (j+ k)}{m}\Bigr)
    + e\Bigl(\frac{i j}{m}\Bigr)z + e\Bigl(\frac{i k}{m}\Bigr) z  \Bigr) .
    \notag
\end{align}
Define polynomial $B_{m,j,k}(z)$ as
\begin{align}
  &B_{m,j,k}(z)
  \notag \\
  &\quad:= (2+2z) A_{m,j,k}(z)
  \notag \\
  &\quad=  \prod_{i=1}^{m}
    \biggl( 1 + e\biggl(\frac{i (j+ k)}{m}\biggr)
    + e\biggl(\frac{i j}{m}\biggr)z + e\biggl(\frac{i k}{m}\biggr) z  \biggr) .
    \label{eq:B-def}
\end{align}
From this definition, $B_{m,j+sm,k+tm}(z) = B_{m,j,k}(z)$ holds for all $s,t\in \mathbb{Z}$.
Hence, if we get $B_{m,j,k}(z)$ for all $j,k \in \intg{m}$, 
we have the Hamming distance enumerator as follows:
\begin{align}
  &\mathcal{D}_{\mathrm{H}}(\mathrm{VT}_{a}(n); z)
  =
  \frac{1}{m^2}
  \sum_{j\in\intg{m}} 
  e \biggl(- \frac{a j}{m}\biggr) F_{m,j}(z),
  \label{eq:dist-VT-mod} \\
  &F_{m,j}(z)
  :=
  \frac{1}{2z+2}\sum_{k \in \intg{m}} B_{m,k,j-k}(z).
  \label{eq:def-F}
\end{align}

This section gives some properties of $B_{m,j,k}(z)$ to calculate the Hamming distance enumerators.
All the proofs are in Appendix.
\begin{lemma} \label{lem:sym}
  For all $m\in \mathbb{Z}^+$, $j,k\in\intg{m}$, the following hold
  \begin{align}
    &B_{m,j,k}(z) = B_{m,k,j}(z),
    \label{eq:diag-sym} \\
    &B_{m,j,k}(z) = (-1)^{j(m+1)} z^m B_{m, m-j, k}\bigl(z^{-1}\bigr),
    \label{eq:ver-sym} \\
    &B_{m,j,k}(z) = (-1)^{k(m+1)} z^m B_{m, j, m-k}\bigl(z^{-1}\bigr).
    \label{eq:hori-sym}
  \end{align}
\end{lemma}
Note that for an $m$th degree polynomial $f(z) = \sum_{i=0}^{m} a_i z^i$,
its reciprocal polynomial is written by $z^m f(z^{-1}) = \sum_{i=0}^{m} a_{m-i}z^{i}$.
Denote the floor function for $x\in\mathbb{R}$, by $\lfloor x \rfloor$, i.e., $\lfloor x \rfloor = \max\{i\in\mathbb{N} \mid i \le x\}$.
By this lemma, we need to derive $B_{m,j,k}(z)$ for only $j,k\in\intg{0, \lfloor \frac{m-1}{2} \rfloor}$.
Moreover, the following lemma allows us to reuse the calculation results.
\begin{lemma} \label{lem:redu1}
  Suppose integers $t, m$ are coprime.
  Then, 
  \begin{equation}
    B_{m, jt, kt}(z) = B_{m, j, k}(z). \label{eq:redu}
  \end{equation}
\end{lemma}

Denote the Chebyshev polynomials of second and third kind, by $U_{n}(z)$ and $V_{n}(z)$, respectively.
The explicit formulas for $U_n(x)$ and $V_n(x)$ are known as
\begin{align*}
  &U_n(x)
  =
  \sum_{k=0}^{\lfloor n/2 \rfloor} (-1)^k \binom{n-k}{k} (2x)^{n-2k}, \\
  &V_n(x)
  =
  \sum_{k=0}^n (-1)^{k} \binom{2n-k}{k} 2^{n-k} (x-1)^{n-k}.
\end{align*}
For some pairs $(j,k)$, we have the explicit formulas of $B_{m,j,k}(z)$.
\begin{lemma} \label{lem:spec}
  Define $d:= \gcd(m,j)$ and denote $m' := m/d$.
  Denote $\bar{m} := \lfloor\frac{m'-1}{2} \rfloor$.
  For all $m\in\mathbb{Z}$ and all $j \in \intg{m}$, the following hold:
  \begin{align}
    &B_{m,j,0}(z)
    = B_{m,0,j}(z)
    = 2^d (1+z)^m \1 [m': \text{odd}], \label{eq:boundary}
    \\
    &B_{m,j,j}(z) \notag \\
    &\quad=
    \begin{cases}
      (-1)^j2^{2d} \bigl(z^2-1\bigr)^d \bigl( U_{ \bar{m}}(z) \bigr)^{2d},
      & (m':\text{even}), \\
      2^{d} (z+1)^d \bigl( V_{ \bar{m} }(z) \bigr)^{2d},
      & (m':\text{odd}), \\
    \end{cases}
    \label{eq:diag} \\
    &B_{m,j,m-j}(z) \notag \\
    &\quad=
    \begin{cases}
      2^{2d} \bigl(z^{2}-1\bigr)^d
      \bigl( z^{\bar{m}} U_{\bar{m}}\bigl(z^{-1}\bigr) \bigr)^{2d},
      & (m':\text{even}),
      \\ %%
      2^{d} (z+1)^d
      \bigl( z^{\bar{m}} V_{\bar{m}}\bigl(z^{-1}\bigr) \bigr)^{2d},
      & (m':\text{odd}) .
    \end{cases}
    \label{eq:anti-diag}
  \end{align}
\end{lemma}

Now, consider the general case of $(j,k)$.
\begin{lemma} \label{lem:redu2}
  Denote $d:=\gcd(m,j,k)$.
  For all $m\in \mathbb{Z}^+$ and $j,k\in\intg{m}$, we have
  \begin{equation*}
    B_{m,j,k}(z)
    =
    \Bigl( B_{\frac{m}{d}, \frac{j}{d}, \frac{k}{d}}(z) \Bigr)^{d} .
  \end{equation*}
\end{lemma}
This lemma implies that $B_{m,j,k}(z)$ is derived from $B_{\frac{m}{d}, \frac{j}{d}, \frac{k}{d}}(z)$, where $\gcd(\frac{m}{d}, \frac{j}{d}, \frac{k}{d}) = 1$.
Hence, the following two lemmas suppose the case of $\gcd(m,j,k) = 1$.
\begin{lemma} \label{lem:odd}
  Suppose $m$ is an odd integer.
  For $j,k\in \intg{m}$ such that $\gcd(m,j,k) = 1$, we get
  \begin{align}
    B_{m,j,k}(z)
    =&~
    2^m   
    (1+z) \notag \\
    &\times\prod_{i=1}^{\frac{m-1}{2}}
    \biggl( \cos \biggl( \pi i\frac{j+k}{m}\biggr)
        + z \cos \biggl( \pi i\frac{j-k}{m}\biggr) \biggr)^2.
    \label{eq:odd}
  \end{align}
\end{lemma}
\begin{lemma} \label{lem:even}
  Suppose $m$ is an even integer.
  For $j,k\in \intg{m}$ such that $\gcd(m,j,k) = 1$, we get
  \begin{align}
    B_{m,j,k}(z)
    =&~
    2^m (1-z^2)  \1[j: \text{odd}] \1[k: \text{odd}] 
    \notag \\
    &\times \prod_{i=1}^{\frac{m}{2}-1}
    \biggl( \cos \biggl( \pi i\frac{j+k}{m} \biggr)
        + z \cos \biggl( \pi i\frac{j-k}{m} \biggr) \biggr)^2.
    \label{eq:even}
  \end{align}
\end{lemma}

\subsection{Algorithms \label{ssec:VT-comp}}

This section shows some algorithms to calculate the Hamming distance enumerators for VT codes.
Firstly, we give a brute-force algorithm (Algorithm \ref{alg:naive}) and evaluate its complexity.
Next, we give an efficient algorithm (Algorithm \ref{alg:FF}) based on the previous section results.

At first, let us consider a brute-force algorithm.
In this algorithm, we enumerate all the codewords in $\mathrm{VT}_a(n)$ and evaluate the Hamming distance between all the pairs of codewords.
Algorithm \ref{alg:naive} gives this brute-force algorithm.
Here, $a\ot b$ represents substituting $b$ for $a$.
This algorithm's complexity is $\mathcal{O}(u^2)$, where $u$ represents the number of codewords in the VT code.
Since the cardinality $u$ of an VT code is approximated by $2^n/(n+1)$, 
this algorithm's complexity is $\mathcal{O}(2^{2n}/n^2)$.
In other words, this brute-force algorithm is exponential time.

\begin{algorithm}[tb]
 %\begin{footnotesize}
   \caption{Brute-force algorithm for calculating $\mathcal{D}_{\mathrm{H}}(\mathrm{VT}_a(n);z)$ \label{alg:naive}}
   \begin{algorithmic}[1]
     \REQUIRE Code length $n$ and an integer $a\in \intg{0,n}$
     \ENSURE Hamming distance distribution $\mathcal{D}_{\mathrm{H}}(\mathrm{VT}_a(n); z) = \sum_{i=0}^{n} D_i z^i$
     \STATE Initialize $D_i \ot 0$ for all $i\in \intg{0,n}$
     \STATE Enumerate codewords $\vect{c}_1, \vect{c}_2, \dots, \vect{c}_u$ in $\mathrm{VT}_a(n)$
     \FOR{$j=1,2,\dots, u$}
     \FOR{$k=1,2,\dots, u$}
     \STATE $D_{\dist_{\mathrm{H}}(\vect{c}_j, \vect{c}_k)} \ot D_{\dist_{\mathrm{H}}(\vect{c}_j, \vect{c}_k)} + 1$
     \ENDFOR
     \ENDFOR
     \STATE Output $\sum_{i=0}^{n} D_i z^i$
   \end{algorithmic}
 %\end{footnotesize}
\end{algorithm}

Next, let us consider an efficient algorithm based on Theorem \ref{the:SC-WT} and lemmas given in the previous section.
This algorithm calculates $\mathcal{D}_{\mathrm{H}}(\mathrm{VT}_a(n);z)$ by deriving $B_{m,j,k}(z)$.
Algorithm \ref{alg:FF} shows the details of this algorithm.
For all $j,k\in\intg{0,n}$, to derive $B_{m,j,k}(z)$ with Lemmas \ref{lem:redu2}, \ref{lem:odd}, and \ref{lem:even},
we need $\mathcal{O}(n^2)$ times multiplication in the real number field.
Hence, the complexity of Step \ref{stp:cal-B} is $\mathcal{O}(n^4)$.
Because the complexity of the other steps are upper bounded by $\mathcal{O}(n^3)$, 
the complexity of this algorithm is $\mathcal{O}(n^4)$.
Thus, Algorithm \ref{alg:FF} has lower complexity than Algorithm \ref{alg:naive}.

\begin{algorithm}[tb]
 %\begin{footnotesize}
   \caption{Algorithm for calculating $\mathcal{D}_{\mathrm{H}}(\mathrm{VT}_a(n);z)$ based on Theorem \ref{the:SC-WT} \label{alg:FF}}
   \begin{algorithmic}[1]
     \REQUIRE Code length $n$ and an integer $a\in \intg{0,n}$
     \ENSURE Hamming distance distribution $\mathcal{D}_{\mathrm{H}}(\mathrm{VT}_a(n); z) $
     \STATE \label{stp:cal-B} Calculate $B_{m,j,k}(z)$ for $j,k\in \intg{0,n}$
     \FOR{$j=0,1,2,\dots,n$}
     \STATE Calculate $F_{m,j}(z)$ by Eq.~\eqref{eq:def-F}
     \ENDFOR
     \STATE Calculate $\mathcal{D}_{\mathrm{H}}(\mathrm{VT}_a(n);z)$ by Eq.~\eqref{eq:dist-VT-mod}
     \STATE Output $\mathcal{D}_{\mathrm{H}}(\mathrm{VT}_a(n);z)$
   \end{algorithmic}
 %\end{footnotesize}
\end{algorithm}

In Step \ref{stp:cal-B} of Algorithm \ref{alg:FF}, to derive $B_{m,j,k}(z)$,
we can reuse the calculation results as in Lemmas \ref{lem:sym}, \ref{lem:redu1}.
By using these results, we summarize an efficient algorithm to derive $B_{m,j,k}(z)$ in Algorithm \ref{alg:comp-B}.

\begin{algorithm}[tb] 
 %\begin{footnotesize}
   \caption{Calculation of $B_{m,j,k}(z)$ \label{alg:comp-B}}
   \begin{algorithmic}[1]
     \REQUIRE Integer $m\in \mathbb{Z}^+$
     \ENSURE Polynomials $B_{m,j,k}(z)$ for $j,k\in \intg{m}$
     \STATE Calculate $B_{m,j,0}(z)$ by Eq.~\eqref{eq:boundary} for $j\in\intg{m}$

     \FOR{$j = 1,2,\dots \lfloor \frac{m-1}{2}\rfloor$}
     \STATE $d \ot \gcd(m,j)$, $d' \ot \gcd(m,j/d)$
     
     \IF{$d' \neq 1$}

     \STATE $B_{m,j,k}(z) \ot B_{m,k,j}(z)$ for $k = 0, 1,\dots j-1$
     \STATE Calculate $B_{m,j,j}(z)$ by Eq.~\eqref{eq:diag}
     \STATE Calculate $B_{m,j,k}(z)$ by Lemmas \ref{lem:redu2}, \ref{lem:odd}, \ref{lem:even} for $k=j+1,\dots, \lfloor \frac{m-1}{2}\rfloor$
     \STATE Set $B_{m,j,k}(z)$ by Eq.~\eqref{eq:ver-sym} for $k=\lfloor \frac{m-1}{2}\rfloor+1,\dots, m-1$
     
     \ELSE
     \STATE $B_{m, j, kj/d} (z) \ot B_{m, d, k}(z)$ for $k \in \intg{m}$
     \ENDIF
     \ENDFOR
     \FOR{$j= \lfloor \frac{m-1}{2}\rfloor+1,\dots, m-1$}
     \STATE Set $B_{m,j,k}(z)$ by Eq.~\eqref{eq:hori-sym} for all $k\in \intg{m}$
     \ENDFOR
     \STATE Output $B_{m,j,k}(z)$ for $j,k\in\intg{m}$
   \end{algorithmic}
 %\end{footnotesize}
\end{algorithm}

%% ADD Example

\subsection{Numerical Example \label{ssec:VT-exam}}
This section gives a numerical example of Hamming distance enumerator for VT codes.

Table \ref{tab:dd_VT15} displays the Hamming distance enumerator for VT codes $\mathrm{VT}_a(15)$ with code length $n=15$.
Note that $\mathcal{D}_{\mathrm{H}}(\mathrm{VT}_a(n);z)$ depends on $d:=\gcd(a,m)$.
In other words, the column labeled with $d=16$ gives the case of $a=0$, and the column labeled with $d=1$ gives the cases of $a=1,3,5,7,9,11,13,15$.

From this table, we see that $D_1 = 0$ for all $d$.
This result is easily checked since the VT codes correct single insertion/deletion.
This table also shows that $D_0$ has the same value.
This value coincides the cardinalities of VT codes \cite{ginzburg1967certain}, \cite{stanley1972study}.

Moreover, by comparing $D_2/ D_0$, we see that the case of $d=4$ has the smallest value.
Hence, in the case of $n=15$, the VT codes with $a=4,12$ are the best codes from the Hamming distance perspective.
On the other hand, in general, the VT code with $a=0$ is the best code from the perspective of the cardinality of the code.
Summarizing above, there are cases that the VT code with $a=0$ is not the best in term of the Hamming distance.

\begin{table}[t]
  \centering
  \caption{Hamming distance enumerator $\mathcal{D}_{\mathrm{H}}(\mathrm{VT}_a(15);z) = \sum_{i=0}^{15} D_iz^i$, where $d:= \gcd(a,m)$ \label{tab:dd_VT15}}
  \begin{tabular}{|r|r|r|r|r|r|} \hline
  $i$ & $d = 16$ & $d = 1$ & $d = 2$ &  $d = 4$  &  $d = 8$ \\ \hline
  0 &      2048 &     2048 &    2048 &     2048  &     2048 \\ \hline
  1 &         0 &        0 &       0 &        0  &        0 \\ \hline
  2 &      7184 &     7168 &    7168 &     7152  &     7184 \\ \hline
  3 &     64496 &    64512 &   64512 &    64528  &    64496 \\ \hline
  4 &    183488 &   183552 &  183456 &   183808  &   183488 \\ \hline
  5 &    375616 &   375552 &  375648 &   375296  &   375616 \\ \hline
  6 &    633152 &   632832 &  633280 &   631616  &   633152 \\ \hline
  7 &    831168 &   831488 &  831040 &   832704  &   831168 \\ \hline
  8 &    828352 &   828736 &  828160 &   832704  &   828352 \\ \hline
  9 &    635968 &   635584 &  636160 &   631616  &   635968 \\ \hline
 10 &    382528 &   382400 &  382624 &   375296  &   382528 \\ \hline
 11 &    176576 &   176704 &  176480 &   183808  &   176576 \\ \hline
 12 &     58384 &    58368 &   58368 &    64528  &    58384 \\ \hline
 13 &     13296 &    13312 &   13312 &     7152  &    13296 \\ \hline
 14 &      2048 &     2048 &    2048 &        0  &     2048 \\ \hline
 15 &         0 &        0 &       0 &     2048  &        0 \\ \hline
  \end{tabular}

\end{table}

\section{Conclusion and Future Works \label{sec:conc}}
This paper has presented the identity of the distance enumerators for the SC codes.
Using this result, we have shown an efficient algorithm to calculate the Hamming distance enumerators for the VT codes.
Moreover, there are cases that $\mathrm{VT}_0(n)$ is not the best in terms of the Hamming distance enumerator.

As future work, we derive the distance enumerator for other SC codes and other distances.

\appendix

\begin{IEEEproof}[Proof of Lemma \ref{lem:sym}]
  From \eqref{eq:B-def}, we get \eqref{eq:diag-sym}.
  Equation \eqref{eq:B-def} leads
  \begin{align*}
    z^m &B_{m, m-j, k}\bigl(z^{-1}\bigr)
    \\  %%
    &=
    \prod_{i=1}^{m} \biggl(
    z
    + e\biggl(\frac{i (-j + k)}{m}\biggr)z
    + e\biggl(\frac{- i j}{m}\biggr) 
    + e\biggl(\frac{i k}{m}\biggr)   
    \biggr)
    \\  &=
    \Biggl(  \prod_{i=1}^{m}  e\biggl(\frac{- i j}{m}\biggr) \Biggr)
    \\
    &\quad\times \prod_{i=1}^{m}
    \biggl(
      e\biggl(\frac{i j}{m}\biggr)z
    + e\biggl(\frac{i k}{m}\biggr)z
    + 1
    + e\biggl(\frac{i (j+k)}{m}\biggr)   
    \biggr)
    \\  &=
    (-1)^{j(m+1)}
    B_{m,j,k}(z).
  \end{align*}
  Hence, we have \eqref{eq:ver-sym}.
  Similarly, we get \eqref{eq:hori-sym}.
\end{IEEEproof}

\begin{IEEEproof}[Proof of Lemma \ref{lem:redu1}]
  For $a\in\mathbb{Z}$, $m\in\mathbb{Z}^+$, denote the remainder in the division of $a$ by $m$, by $\langle a \rangle_{m}$.
  Then, since $t$ satisfies $\gcd(t,m) = 1$, we have $\{ \langle it \rangle_{m} \mid i\in\intg{m} \} = \intg{m}$.
  Combining this and \eqref{eq:B-def}, we get \eqref{eq:redu}.
\end{IEEEproof}

\begin{IEEEproof}[Proof of Lemma \ref{lem:spec}]
  The first equality of \eqref{eq:boundary} follows \eqref{eq:diag-sym}.
  From \eqref{eq:B-def}, we get
  \begin{align}
    B_{m,j,0}(z)
    =
    (1+z)^m
    \prod_{i=1}^{m}
    \biggl(  1 + e\biggl(\frac{i j}{m}\biggr) \biggr).
    \label{eq:der-B-1}
  \end{align}
  Note that
  \begin{align*}
    \prod_{i=1}^{m}
    \biggl( 1 + e\biggl(\frac{i j}{m}\biggr)u  \biggr)
    =
    \bigl( 1 - (-u)^{m/d} \bigr)^{d}.
  \end{align*}
  Substituting $u=1$, we get 
  \begin{align}
    \prod_{i=1}^{m}
    \biggl(  1 + e\biggl(\frac{i j}{m}\biggr) \biggr)
    =
    2^d \1 [m/d : \text{odd}] . \label{eq:der-B-2}
  \end{align}
  Combining \eqref{eq:der-B-1} and \eqref{eq:der-B-2}, we get the second equality of \eqref{eq:boundary}.

  From \eqref{eq:B-def}, we have
  \begin{align}
    B_{m,j,j}(z) 
    &=
    \prod_{i=1}^{m} \biggl(
    1
    +  e\biggl(\frac{2i j}{m}\biggr)
    + 2e\biggl(\frac{ i j}{m}\biggr) z
    \biggr)
    \notag \\ &=
    \prod_{i=1}^{m}  \biggl( 2e\biggl(\frac{i j}{m}\biggr) \biggr)
    \prod_{i=1}^{m}
    \biggl(
    \cos \biggl(2\pi \frac{ij}{m} \biggr)
    +  z
    \biggr)
    \notag \\ &=
    2^m (-1)^{j(m+1)}
    \prod_{i=1}^{m}
    \biggl( \cos \biggl(2\pi \frac{i j/d}{m/d} \biggr) +  z \biggr)
    \notag \\ &=
    (-1)^{j(m+1)}
    \Biggl\{
      2^{m'}
      \prod_{i=1}^{m'}
      \biggl( z + \cos \biggl(\frac{ 2\pi i }{m'} \biggr) \biggr)
    \Biggr\}^d
    \notag \\ &=:
    (-1)^{j(m+1)}
    \bigl\{ G_{m'}(z) \bigr\}^d, \label{eq:der-diag-1}
  \end{align}
  where the fourth equality follows from $\gcd(j/d, m') = 1$.
  The Chebyshev polynomials of second and third kind (e.g., see \cite{mason2002chebyshev}) are expressed as
  \begin{align*}
    U_n(z)
    &=
    2^{n} \prod_{i=1}^n\biggl(z - \cos \biggl( \pi \frac{i}{n+1}  \biggr)  \biggr)
    \\
    &=
    2^{n} \prod_{i=1}^n\biggl(z + \cos \biggl( \pi \frac{i}{n+1}  \biggr)  \biggr),
    \\ %%
    V_n(z)
    &=
    2^n  \prod_{i=1}^n\biggl(z - \cos \biggl( \pi \frac{2i-1}{2n+1} \biggr) \biggr)
    \\ &=
    2^n  \prod_{i=1}^n\biggl(z + \cos \biggl( \pi \frac{2i}{2n+1} \biggr) \biggr).
  \end{align*}
  Consider even $m'$, i.e., $m' = 2\bar{m} + 2$.
  Then,
  \begin{align}
    G_{m'}(z)
    &=
    4 (z^2-1)
    \Biggl\{
      2^{\bar{m}}
      \prod_{i=1}^{\bar{m}}
      \biggl( z + \cos \biggl(\pi \frac{i }{\bar{m}+1} \biggr)  \biggr)
    \Biggr\}^2
    \notag \\ &=
    4 (z^2-1) \{U_{\bar{m}}(z)\}^2. \label{eq:der-diag-2}
  \end{align}
  Here, 
  $\prod_{i=\bar{m}+2}^{2\bar{m}+1}
   \bigl( z + \cos \bigl( \frac{\pi i }{\bar{m}+1} \bigr) \bigr)
   =
   \prod_{i=1}^{\bar{m}}
   \bigl(  z + \cos \bigl( \frac{\pi i }{\bar{m}+1} \bigr) \bigr)$
  leads the first equality.
  Consider odd $m'$, i.e., $m' = 2\bar{m}+1$.
  Then,
  \begin{align}
    G_{m'}(z)
    &=
    2(z+1)
    \Biggl\{
      2^{\bar{m}}  \prod_{i=1}^{\bar{m}}
      \biggl( z + \cos \biggl(\pi \frac{2i }{2\bar{m}+1} \biggr) \biggr)
    \Biggr\}^2
    \notag \\ &=
    2(z+1) \bigl\{ V_{\bar{m}}(z) \bigr\}^2.
    \label{eq:der-diag-3}
  \end{align}
  Here, 
  $\prod_{i=\bar{m}+2}^{2\bar{m}+1}
  \bigl( z + \cos \bigl(\pi \frac{2i }{2\bar{m}+1} \bigr) \bigr)
  =
   \prod_{i=1}^{\bar{m}}
   \bigl(  z + \cos \bigl(\pi \frac{2i }{2\bar{m}+1} \bigr) \bigr)$
  leads the first equality.
  Combining \eqref{eq:der-diag-1}, \eqref{eq:der-diag-2}, and \eqref{eq:der-diag-3}, we get \eqref{eq:diag}.

  Equations \eqref{eq:diag} and \eqref{eq:hori-sym} lead \eqref{eq:anti-diag}.
\end{IEEEproof}

\begin{IEEEproof}[Proof of Lemma \ref{lem:redu2}]
  Equation \eqref{eq:B-def} gives the lemma.
\end{IEEEproof}

From \eqref{eq:B-def}, we get
\begin{align}
  B_{m,j,k}&(z)
  \notag \\ =&~
  \prod_{i=1}^{m} \biggl(
  1 + e \biggl( i\frac{j+k}{m} \biggr)
  + e \biggl( \frac{ij}{m} \biggr) z  + e \biggl( \frac{ik}{m} \biggr) z  \biggr)
  \notag \\ =&~
  \prod_{i=1}^{m} e\biggl( i\frac{j+k}{2m}  \biggr)
  \biggl(
  e \biggl( - i\frac{j+k}{2m} \biggr) + e \biggl( i\frac{j+k}{2m} \biggr) 
  \notag \\ &\hspace{25mm} + e \biggl( i\frac{j-k}{2m} \biggr) z  + e \biggl( i\frac{k-j}{2m}i \biggr) z  \biggr)
  \notag \\ =&~
  2^m (-1)^{(j+k)(m+1)/2}
  \notag \\  &\times
  \prod_{i=1}^{m}
  \biggl( \cos \biggl( \pi i\frac{j+k}{m} \biggr)
      + z \cos \biggl( \pi i\frac{j-k}{m} \biggr) \biggr).
  \label{eq:B-cos}
\end{align}
Based on this equation, we will derive Lemmas \ref{lem:odd} and \ref{lem:even}.

\begin{IEEEproof}[Proof of Lemma \ref{lem:odd}]
  From \eqref{eq:B-cos}, we have
  \begin{align*}
    &B_{m,j,k}(z)
    \\ &~~=
    2^m   (-1)^{(j+k)(m+1)/2}  (1+z)
    \\ &\qquad\times
    \prod_{i=1}^{\frac{m-1}{2}}
    \biggl( \cos \biggl( \pi i\frac{j+k}{m} \biggr)
        + z \cos \biggl( \pi i\frac{j-k}{m} \biggr) \biggr)
    \\ &\qquad\times
    \prod_{i=\frac{m-1}{2}+1}^{m-1}
    \biggl( \cos \biggl( \pi i\frac{j+k}{m} \biggr)
        + z \cos \biggl( \pi i\frac{j-k}{m} \biggr) \biggr)
    \\ &~~=
    2^m (-1)^{(j+k)(m+1)/2}  (1+z)
    \\ &\qquad\times\prod_{i=1}^{\frac{m-1}{2}}  (-1)^{j+k}
    \biggl( \cos \biggl( \pi i\frac{j+k}{m} \biggr)
        + z \cos \biggl( \pi i\frac{j-k}{m} \biggr) \biggr)^2
    \\ &~~=
    2^m (1+z)
    \prod_{i=1}^{\frac{m-1}{2}}
    \biggl( \cos \biggl( \pi i\frac{j+k}{m} \biggr)
        + z \cos \biggl( \pi i\frac{j-k}{m} \biggr) \biggr)^2.
  \end{align*}
  This concludes the proof.
\end{IEEEproof}

\begin{IEEEproof}[Proof of Lemma \ref{lem:even}]
  Let us start from \eqref{eq:B-cos}.
  Note that
  \begin{align*}
    &\biggl(
    \cos \biggl(\pi \frac{m}{2}\frac{j+k}{m}  \biggr)
    + z   \cos \biggl( \pi \frac{m}{2}\frac{j-k}{m} \biggr) \biggr)
    \\ &\quad=
    \1[j+k : \text{even}]
    \bigl\{ (-1)^{(j+k)/2} + z (-1)^{(j-k)/2} \bigr\}
    \\ &\quad=
    \1[j : \text{odd}]    \1[k : \text{odd}]
    (-1)^{(j+k)/2} \bigl\{ 1  + z (-1)^{-k} \bigr\}
    \\ &\quad=
    \1[j : \text{odd}]    \1[k : \text{odd}]
    (-1)^{(j+k)/2} ( 1 - z ).
  \end{align*}
  Here, the first equality follows from the fact that $(j+k)$ is even iff $(j-k)$ is even;
  The second equality follows from even $m$ and $\gcd(m,j,k) = 1$. 
  Combining this equation and \eqref{eq:B-cos} leads
  \begin{align*}
    B_{m,j,k}&(z)
    \\ =&~
    2^m   (-1)^{(j+k)/2}  (1+z)
    \\
    &\times \1[j : \text{odd}]    \1[k : \text{odd}]
    (-1)^{(j+k)/2}( 1 - z )
    \\ &\times
    \prod_{i=1}^{\frac{m}{2}-1}
    \biggl( \cos \biggl( \pi i \frac{j+k}{m} \biggr)
        + z \cos \biggl( \pi i \frac{j-k}{m} \biggr) \biggr)
    \\ &\times
    \prod_{i=\frac{m}{2}+1}^{m-1}
    \biggl( \cos \biggl( \pi i \frac{j+k}{m} \biggr)
        + z \cos \biggl( \pi i \frac{j-k}{m} \biggr) \biggr)
    \\ =&~
    2^m  (1-z^2)  \1[j : \text{odd}]    \1[k : \text{odd}]
    \\ &\times \prod_{i=1}^{\frac{m}{2}-1}
    \biggl( \cos \biggl( \pi i \frac{j+k}{m} \biggr)
        + z \cos \biggl( \pi i \frac{j-k}{m} \biggr) \biggr)^2.
  \end{align*}
  This concludes the proof.
\end{IEEEproof}

\section*{Acknowledgment}

This work is supported by Inamori Research Grants.


\end{document}